\documentclass[12pt,epsf]{article}
\usepackage{graphicx,color}
\begin{document}

\newcommand{\gsim}{ \mathop{}_{\textstyle \sim}^{\textstyle >} }
\newcommand{\lsim}{ \mathop{}_{\textstyle \sim}^{\textstyle <} }

\baselineskip 0.7cm

\begin{titlepage}

\begin{flushright}
UT-06-01
\end{flushright}

\vskip 1.35cm
\begin{center}
{\large \bf
Minimal Supergravity, Inflation, and All That
}
\vskip 1.2cm
M.~Ibe${}^{1}$, Izawa~K.-I.${}^{1,2}$, Y.~Shinbara${}^{1}$, and T.T.~Yanagida${}^{1,2}$
\vskip 0.4cm

${}^1${\it Department of Physics, University of Tokyo,\\
     Tokyo 113-0033, Japan}

${}^2${\it Research Center for the Early Universe, University of Tokyo,\\
     Tokyo 113-0033, Japan}

\vskip 1.5cm

\abstract{
We consider an inflationary model in the
hidden-sector broken supergravity with an effectively large cutoff.
The inflaton decay into right-handed neutrinos naturally
causes the observed baryon asymmetry of the universe
with a reheating temperature low enough to avoid the gravitino
overproduction.
We emphasize that all the
phenomenological requirements from cosmology and particle physics are
satisfied in the large-cutoff theory.
}
\end{center}
\end{titlepage}

\setcounter{page}{2}

\section{Introduction}

The landscape of many vacua%
\footnote{
The vacua here have extended meaning
which indicates the backgrounds in the theory (moduli)
space, or the landscape.
}
is a plausible structure in the fundamental
theory of physical laws in nature.
In particular, this structure is expected
as one of the theoretical ingredients
to understand the observed small cosmological constant
\cite{Wei}.
However, the anthropically allowed region of vacua
in the landscape seems too large to be predictive enough
in the presence of a variety of couplings.
Thus, it is a challenging problem to derive
further physical consequences from the landscape of vacua.

The (non-)presence of inflationary dynamics
is a promising candidate as the first criterion to select realistic vacua
\cite{Iza}.
We can naturally expect that macroscopic universe is realized through
inflation from fundamental-scale physics.
Moreover, under the dynamics of inflation, mediocrity principle
\cite{Vil}
may prefer long-lasting inflations which result in
larger-volume universes where more habitable galaxies are
produced.
In this respect, multiple inflations
\cite{Kaw}
give a remarkable possibility to be considered
\cite{Ibe}.

In a recent article
\cite{Ibe},
we have pointed out that the
inflationary dynamics possess a potential
to select minimal supergravity as a large-cutoff theory,
where the gravitational scale $M_G$ is smaller than the
cutoff scale $M_*$ stemming from the fundamental theory.
Such a large-cutoff supergravity naturally causes
multiple slow-roll inflations, which possibly meet mediocrity principle.

The large-cutoff theory is also attractive from the viewpoint of
particle-physics phenomenology: First of all, the suppression of the
flavor-changing neutral currents is automatic in the large-cutoff
theory, since all of the higher-dimensional operators are suppressed by 
the large cutoff $M_*$ except for the genuine gravitational interactions.
In detail,
the large-cutoff supergravity predicts a hierarchical spectrum
\cite{Ibe}
of supersymmetric
particles as $m_0 \gg |M_i|$, where $m_0$ is the universal soft mass for
sfermions and $M_i$ the gaugino masses ($i=1,2,3$).
Thus, the current chargino mass bound suggests heavy sfermions at several TeV.
Such a soft mass parameter belongs to the parabolic
\cite{Ibe}
or hyperbolic
\cite{Cha}
regime allowed for a given $\mu$ parameter.
Indeed the recent detailed analysis
\cite{Ibe}
has confirmed that the region with large sfermion masses
along the small-$\mu$-parameter curve continued from the focus point
\cite{Fen}
is consistent with the electroweak symmetry breaking
\cite{Bar}.
In the region of
heavy sfermions and light gauginos (an order of magnitude lighter than
the sfermions), the constraint
from CP violation is rather weak and even order one CP-violating phases are
allowed for $m_0 \gsim 10$TeV
\cite{Olive:2005ru}.
Furthermore, the lightest supersymmetric particle
can explain the dark matter density of the present universe in the above
mass region for supersymmetric particles
\cite{Mor}.

In this paper, we discuss a minimal new inflation model
\footnote{We suspect that multiple stages of inflation imply
that the primordial inflation at the last stage tends to be
a new inflation, since it seems naturally
realized with a lower energy scale than that of other types of inflation.
The discussion section includes comments on the case of other inflations.
}
as an example in the framework of the
large-cutoff supergravity with emphasis on baryon asymmetry
generated by leptogenesis
\cite{Fuk}
to complete a model of the large-cutoff hypothesis.\footnote{There are
other new inflation models in the framework of
supergravity\cite{Asaka:1999jb}, although these inflation models can not
explain the observed spectral index in the large-cutoff hypothesis.}
We claim that all the phenomenological requirements from
cosmology and particle physics are satisfied in a certain parameter
region of the large-cutoff theory.

\section{Supergravity new inflation}

We adopt a new inflation model considered in
Ref.\cite{Yan, Izaw}.
As an effective field theory for
an inflaton chiral superfield $\tilde \phi$,
the superpotential is given by
\begin{equation}
 \begin{array}{rr}
 W={\tilde v}^2 {\tilde \phi} -\frac{\tilde g}{n+1}{\tilde \phi}^{n+1},
 &
 \label{inflaton-tilde}
 \end{array}
\end{equation}
for $n \ge 3$ and the K\"ahler potential is given by
\begin{equation}
 K = {\tilde K}\left| {\tilde \phi} \right|^2
 + \frac{\tilde k}{4}\left| {\tilde \phi} \right|^4 + \cdots,
\end{equation}
where we have taken the unit with the reduced Planck scale
$M_{G} \simeq 2.4 \times 10^{18}$GeV equal to
one. The positive parameters $\tilde K$, $\tilde g$, and $\tilde k$
are of orders $1$, $10^{-(n-2)}$, and $10^{-2}$, respectively,
for our large-cutoff hypothesis $M_* \simeq 10 M_G$.
The tiny scale ${\tilde v}^2 >0$ can be generated dynamically
\cite{Hot} and the ellipsis denotes higher-dimensional operators which
may be neglected in the following analysis.

For the canonically normalized field $\phi = \sqrt{\tilde K} {\tilde \phi}$,
the superpotential is given by
\cite{Kumekawa:1994gx}
\begin{equation}
 \begin{array}{rr}
 W=v^2 \phi -\frac{g}{n+1}\phi^{n+1}, &
 \label{inflaton}
 \end{array}
\end{equation}
and the K\"ahler potential is given by
\begin{equation}
 K = \left| \phi \right|^2 + \frac{k}{4}\left| \phi \right|^4 + \cdots,
\end{equation}
where we have defined
\begin{equation}
 {\tilde v}^2=v^2 \sqrt{\tilde K}, \quad
 {\tilde g}=g {\tilde K}^{n+1 \over 2}, \quad
 {\tilde k}=k {\tilde K}^2.
 \label{canonical}
\end{equation}

The effective potential for the lowest component of $\phi$ is given by
\begin{equation}
 V=e^K \left\{ \left(\frac{\partial ^2 K}
 {\partial \phi \partial \phi^\dagger}\right)^{-1}
 \left|D W \right|^2 -3\left|W \right|^2\right\},\label{sugpot}
\end{equation}   
where 
\begin{equation}
 DW = \frac{\partial W}{\partial \phi}
 + \frac{\partial K}{\partial \phi}W.
\end{equation}
Thus, the potential of the inflaton field
$\varphi=\sqrt{2} \mathrm{Re} \ \phi$ is
approximately given by
\begin{equation}
 V(\varphi) \simeq v^4 - \frac{k}{2}v^4 \varphi^2
 -\frac{g}{2^{\frac{n}{2}-1}}v^2\varphi^n
 +\frac{g^2}{2^n}\varphi^{2n}
\label{potential}
\end{equation}
for the inflationary period near the origin $\varphi > 0$.

The inflationary regime is determined by the slow-roll condition
\cite{Lyt}
\begin{equation}
 \epsilon(\varphi) = \frac{1}{2}\left(\frac{V^\prime(\varphi)}
 {V(\varphi)}\right)^2 \le 1,\quad \left|\eta(\varphi)\right| \le 1,
 \label{slowroll}
\end{equation}
where 
\begin{equation}
 \eta(\varphi)=\frac{V^{\prime\prime}(\varphi)}{V(\varphi)}.
\end{equation}
For the potential Eq.(\ref{potential}), we obtain
\begin{eqnarray}
 \epsilon(\varphi) \simeq \frac{1}{2}
 \left(\frac{-k v^4 \varphi-\frac{gn}
 {2^{\frac{n}{2}-1}}v^2\varphi^{n-1}}{v^4}\right)^2, \\
 \eta(\varphi) \simeq \frac{-kv^4 -\frac{g}
 {2^{\frac{n}{2}-1}}n(n-1)v^2\varphi^{n-2}}{v^4}.
\end{eqnarray}
The slow-roll condition Eq.(\ref{slowroll}) is satisfied for
$\varphi\le\varphi_f$ where
\begin{equation}
 \varphi_f \simeq \sqrt{2}\left(\frac{(1-k)v^2}
 {gn(n-1)}\right)^{\frac{1}{n-2}},
\end{equation}
which yields the value of the inflaton field at the end of inflation.

\begin{figure}
 \begin{center}
 \includegraphics[width=7.5cm]{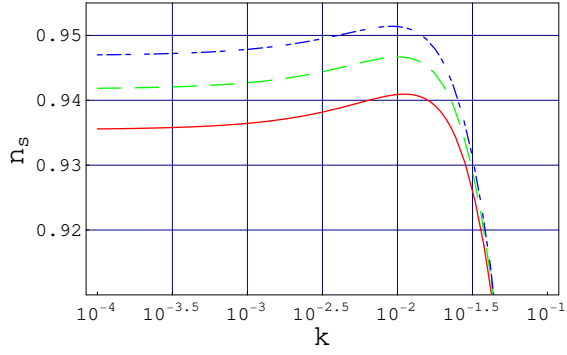}
 \caption{The $k$ dependence of the spectral index $n_s$ for $n=4$. The red
 (solid) line corresponds to the $e$-fold number $N_e$=45, the green (dashed)
 line to $N_e$=50, and the blue (dash-dotted)
 line to $N_e$=55. For $k=0$, $n_s \simeq 1-6/(2N_e+3)$.}
 \label{spectrum1}
 \end{center}
\end{figure}

The value $\varphi_{N_e}$ of the inflaton corresponding to the $e$-fold number $N_e$ is
given by
\begin{eqnarray}
 N_e \simeq \int_{\varphi_f}^{\varphi_{N_e}} d\varphi
 \frac{V(\varphi)}{V^\prime(\varphi)}
   \simeq \int_{\varphi_f}^{\varphi_{N_e}}
 d\varphi \frac{v^4}{-kv^4\varphi-\frac{gn}
 {2^{\frac{n}{2}-1}}v^2\varphi^{n-1}}.
\end{eqnarray}
This leads to
\begin{equation}
 \varphi_{N_e}^{n-2} \simeq \frac{kv^2 2^{\frac{n}{2}-1}}{gn}
 \left\{\frac{1+k(n-2)}{1-k}e^{{N_e}k(n-2)}-1\right\}^{-1}. 
\end{equation}
Hence the spectral index of the density fluctuations is given by
\cite{Izaw}
\begin{eqnarray}
 n_s &\simeq& 1-6\epsilon(\varphi_{N_e}) + 2 \eta(\varphi_{N_e})\\
  &\simeq& 1 - 2 k \left[1 + \frac{n-1}
 {\left\{1+\frac{k}{1-k}(n-1)\right\}e^{{N_e}k(n-2)}-1}\right].
 \label{spectrum}
\end{eqnarray}
Note that the spectral index does not depend on $v^2$ and $g$
explicitly. We show the $k$ dependence of the spectral index
$n_s$ for $n=4$ and $N_e=45, 50$, $55$
\cite{Yan,Izaw}
in Fig.\ref{spectrum1}.

Now we proceed to determine the inflation scale $v$ from the density
fluctuations. The amplitude of primordial density fluctuations is given by
\begin{eqnarray}
 \frac{\delta \rho}{\rho} \simeq \frac{1}
 {5\sqrt{3}\pi}\frac{V^{3 \over 2}(\varphi_{N_0})}{|V^\prime(\varphi_{N_0})|}
                          \simeq \frac{1}{5\sqrt{3}\pi}\frac{v^6}
 {kv^4\varphi_{N_0} + \frac{gnv^2}{2^{\frac{n}{2}-1}}\varphi_{N_0}^{n-1}},
\end{eqnarray}
where $\varphi_{N_0}$ is the value of inflaton field at the epoch of the
present-horizon exit.
Thus we obtain
\begin{eqnarray}
 v^{\frac{2n-6}{n-2}} \simeq \sqrt{2}\frac{V^{3 \over 2}(\varphi_{N_0})}{|V^\prime(\varphi_{N_0})|}\left[\frac{k}{gn}\left\{\frac{1+k(n-2)}{1-k}e^{N_0k(n-2)}-1\right\}^{-1}\right]^{\frac{1}{n-2}}\cr
\quad \times\left[k+k\left\{\frac{1+k(n-2)}{1-k}e^{N_0k(n-2)}-1\right\}^{-1}\right].
\label{v}
\end{eqnarray}
Owing to the COBE normalization
\begin{equation}
 \frac{V^{3 \over 2}(\varphi_{N_0})}{|V^\prime(\varphi_{N_0})|} \simeq 5.3 \times 10^{-4},
\end{equation}
the scale $v$ is expressed as
\begin{equation}
 v \simeq 10^{12} \mathrm{GeV} \times \mathcal{C}(k,N_0) \times \left(\frac{0.1}{g}\right)^{1/2}\label{approv},
\end{equation}
for $n=4$, $N_0 \simeq 50$, and $k\lsim0.01$, where $\mathcal{C}(k,N_0)$ is
a function of order unity.

\begin{figure}
 \begin{center}
 \includegraphics[width=7.5cm]{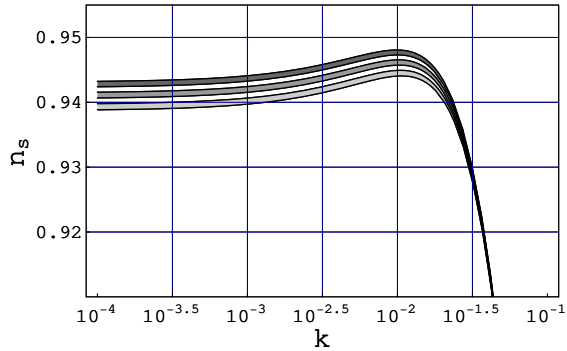}
 \caption{The $k$ dependence of the spectral index $n_s$
 for $n=4$. The shaded regions correspond to
  $T_R=10^5,10^7,10^9$GeV from below, and the lower lines for $g=0.1$ and
  the upper lines for $g=0.01$.}
 \label{spectrum3}
 \end{center}
\end{figure}

On the other hand,
the $e$-fold number of the present horizon is also given by
\begin{equation}
 N_0\simeq 67+\frac{1}{3}\ln H +\frac{1}{3}\ln T_{R} \simeq 67
 +\frac{1}{3}\ln \frac{v^2}{\sqrt{3}} + \frac{1}{3}\ln T_R,
\label{efold}
\end{equation}
where $H$ denotes the Hubble scale at the horizon exit
and $T_R$ the reheating temperature.
By means of Eq.(\ref{spectrum}), (\ref{v}), and (\ref{efold}), we can
determine $v$ and $N_0$ from $g$, $k$, and $T_R$. For $n=4$, $g\sim 0.1$,
$k\sim0.01$, and $T_R \sim 10^{5-9}$GeV, the inflation scale $v$ is
given by ${\cal O}(10^{12})$GeV, and the $e$-fold number $N_0$ of the present
horizon is given by $47.6-50.6$.
In Fig.\ref{spectrum3}, we show the $k$ dependence of the spectral index $n_s$
for the reheating temperature $T_R=10^5,10^7$, and $10^9$ GeV.
We conclude that the implication
${\tilde g} \sim {\tilde k} \sim 0.01$
of the large-cutoff hypothesis
\footnote{For instance, Eq.(\ref{canonical}) yields $g=0.1$ and $k=0.01$
for ${\tilde K}=0.5$, ${\tilde g}=0.018$
and ${\tilde k}=0.0025$.}
is consistent with an experimental value $n_s=0.95 \pm 0.02$
\cite{BOOMERANG} of the spectral index for a wide range of the reheating
temperature.\footnote{
The inflaton as a massless scalar field in
the de Sitter background
has quantum fluctuations whose amplitude is given by $\Delta \varphi \sim
H/(2 \pi)$.
Thus the amplitude $\Delta \varphi$ at $\varphi=\varphi_{N_0}$ is given by
\begin{eqnarray}
 \Delta \varphi|_{\varphi_{N_0}} \sim \frac{\sqrt{2\epsilon(\varphi_{N_0})}}{2\pi\sqrt{3}} \frac{V(\varphi_{N_0})^{3 \over 2}}{V'(\varphi_{N_0})}
            \simeq \frac{1}{2\pi\sqrt{3}}\frac{V(\varphi_{N_0})^{3 \over 2}}{V'(\varphi_{N_0})}\left(k+\frac{gn}{2^{\frac{n}{2}-1}}\frac{\varphi_{N_0}^{n-2}}{v^2}\right)\varphi_{N_0}.
\nonumber
\end{eqnarray}
For $n=4, g \sim0.1, k \sim 0.01$, and $T_R\sim 10^{5-9}$GeV,
the fluctuation amplitude $\Delta
\varphi|_{\varphi_{N_0}}$ takes a value of order $10^{-6} \varphi_{N_0}$,
which is much less than the mean-field value $\varphi_{N_0}$ to justify
the above slow-roll analysis.}

\section{The gravitino mass}

In the previous section, we have confirmed that
the new inflation model in the large-cutoff hypothesis
is consistent with the cosmological observations.
In this section, we discuss
the gravitino problem under such an inflationary scenario.  

As considered in
Ref.\cite{Yan},
we assume that the positive energy  $\Lambda^4_{\rm{SUSY}}$
of the SUSY breaking
is dominantly canceled out by the negative energy at the inflaton potential minimum.
Namely we impose
\begin{equation}
 \Lambda^4_{\rm{SUSY}}-3|W(\phi_0)|^2=0,
\end{equation}
where $\phi_0$ is the minimum point of $\phi$ in Eq.(\ref{sugpot}). 

Then we obtain the gravitino mass as
\begin{equation}
 m_{3/2} \simeq \frac{\Lambda^2_{\rm{SUSY}}}{\sqrt{3}}=W(\phi_0).
\end{equation}
The value of $\phi_0$ is approximately given by
\begin{equation}
 \phi_0 \simeq \left(\frac{v^2}{g}\right)^{\frac{1}{n}}.
 \label{phivev}
\end{equation}
Consequently the gravitino mass is given by
\begin{equation}
 m_{3/2} \simeq  \frac{n v^2}{n+1}\left(\frac{v^2}{g}\right)^{\frac{1}{n}}
         \simeq  9 \mathrm{TeV} \times \left(\frac{0.1}{g}\right)^{3 \over 2}.
 \label{gmass}
\end{equation}
The second equality holds for $n=4$,
where we have used Eq.(\ref{approv}) and omitted the
weak dependence on $k$ and $T_R$.

\begin{figure}
 \begin{center}
 \includegraphics[width=7.5cm]{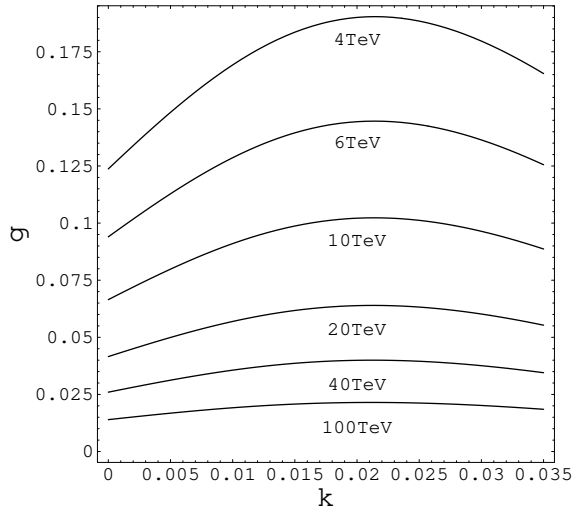}
 \caption{The contours of the gravitino mass for $n=4$
 and $T_R=4 \times 10^6$GeV. The dependence on
 the reheating temperature is very weak.}
 \label{gravitino}
 \end{center}
\end{figure}

More precisely, by means of Eq.(\ref{v}) and (\ref{efold}),
the gravitino mass can be
expressed as a function of $g$, $k$, and $T_R$,
although the dependence on
$T_R$ is very weak, as can be seen from Eq.(\ref{v}) and (\ref{efold}).
The result is shown in
Fig.\ref{gravitino}.
For $g \lsim 0.2$ and $k \lsim 0.035$,
the gravitino mass is larger than $4$TeV,
which may avoid the gravitino overproduction
for a reheating temperature $T_R \sim 10^{6-7}$GeV
\cite{Koh}.

In contrast,
the sfermion soft mass is given as $m_0 \simeq m_{3/2}$ if no $D$-term contributes to the SUSY breaking.
Thus, $m_0 < 10$TeV implies $g \gsim 0.07$ for $k \lsim 0.035$.

\section{Reheating for baryogenesis}
\label{se:baryon}
Now we are ready to consider the baryon asymmetry
in the present new inflation model with the large cutoff.

We assume the baryon asymmetry is generated by leptogenesis
\cite{Fuk}
through non-thermal production of
right-handed neutrinos, as investigated in
Ref.\cite{Mor,Fuk},
which provides a numerical estimate
\begin{equation}
 \frac{n_B}{s} \simeq 8.2 \times 10^{-11} \left(\frac{T_R}{10^6\mathrm{GeV}}\right)\left(\frac{2m_N}{m_\phi}\right)\left(\frac{m_{\nu_3}}{{0.05\mathrm{eV}}}\right)\frac{1}{\sin^2 \beta}\delta_{eff}.
 \label{lepto}
\end{equation}
Here $m_N$, $m_\phi$, and $m_{\nu_3}$ are the masses of the right-handed
neutrino $N$, the inflaton $\phi$ and the heaviest (active) neutrino, respectively.
The phase $\delta_{eff}$ is the effective CP phase defined in
Ref.\cite{Fuk} and $\tan \beta$ is the ratio of the vacuum
expectation value of up- and down-type Higgs bosons in the MSSM.
The reheating temperature is given by  
\begin{equation}
 T_R \simeq \left(\frac{10}{g_*\pi^2} \Gamma_\phi^2\right)^{\frac{1}{4}},
\label{tr}
\end{equation}
where $\Gamma_\phi$ is the decay width of the inflaton and $g_*$ is the effective number of massless degrees of
freedom to be taken as $g_*=228.75$ numerically.
Note that the inflaton mass
\begin{equation}
 m_\phi \simeq n v^2 \left({v^2 \over g}\right)^{-{1 \over n}}
\label{mphi}
\end{equation}
in our new inflation model
also weakly depends on the $k$ and the reheating
temperature $T_R$, as is the case for the gravitino mass in
Eq.(\ref{gmass}).

Let us introduce the following superpotential interaction
as the dominant source of the $N$ production:
\footnote{The inflaton
field is also expected to decay through couplings
with light fields $\psi_i$ in the K\"ahler potential
such as $\sum_i c_i|\phi|^2 |\psi_i|^2$.
However, the decay width $\Gamma \sim \sum_i |c_i|^2\phi_0^2m_\phi^3$
through these couplings is so small that
we neglect such contributions.}
\begin{equation}
 \delta W = \frac{h}{2(n-1)}\phi^{n-1}N^2,
 \label{hNN}
\end{equation}
where $h$ is a positive parameter of the order of the inflaton
self-coupling $g$. 
\footnote{Here, we assign the same charge for $\phi$ and
$N$ under $Z_{2n}$ R-symmetry, while we
assign the matter parity $+$ for
$\phi$ and $-$ for $N$.
Hence we expect the presence of
such operators as $\phi^{n-3}N^{4}$ in addition to Eq.(\ref{hNN}).
We do not include such operators since
the operator Eq.(\ref{hNN}) with the smallest number of $N$
dominates the reheating and leptogenesis.}
The coupling Eq.(\ref{hNN}) gives a decay width
\begin{equation}
  \Gamma_\phi \simeq \frac{|h|^2}{16\pi}\phi_0^{2(n-2)}m_\phi.
\label{gammaphi}
\end{equation}
From this decay width the reheating temperature after inflation for
$n=4$ is given by
\footnote{The cross term between $\phi^{n-1}N^2$ and
$v^2\phi$ in the superpotential
gives a comparable decay width.
We neglect this contribution since it does not essentially affect
our conclusions.}
\begin{equation}
T_R \simeq 2.6\times 10^6 \mathrm{GeV} \left(\frac{h}{0.1}\right)\left(\frac{0.1}{g}\right)^{5/4},
\end{equation}
where we have omitted the weak dependence on $k$ in Eq.(\ref{approv}).  
Therefore, the reheating temperature $T_R \sim 10^{6-7}$ GeV is typical
in this model. As mentioned above, this reheating temperature is low
enough to avoid the gravitino overproduction.

Note that the operator Eq.(\ref{hNN}) also gives the
Majorana mass to the neutrino:
\begin{equation}
 m_N=\frac{h}{n-1} \phi_0^{n-1}\simeq \frac{h}{n-1}\left(\frac{v^2}{g}\right)^{1-\frac{1}{n}}.
\end{equation} 
Thus the mass inequality $2m_N<m_\phi$, namely,
\begin{equation}
 h <\frac{n(n-1)}{2}g,
\label{tiny}
\end{equation} 
is satisfied with a typical parameter set $g \sim h$.
This is appropriate for the non-thermal production of neutrinos
which leads to the non-thermal leptogenesis.

Based on the above setup, we now
estimate the baryon asymmetry due to the decay of inflaton
\footnote{
In our setup, we also have an additional contribution to the baryon asymmetry and the gravitino abundance.
However, as we see in Appendix, this contribution
is small in typical parameter region so that we neglect this
contribution in the following analysis.}
in our model as a function of the couplings $g,k$, and the
reheating temperature $T_R$.
The baryon asymmetry is determined
by four independent parameters $g,k,v$, and $h$. 
In terms of the observed density fluctuations,
we can represent $v$ with the other parameters.
We further use the reheating temperature as an input parameter
instead of $h$ by means of Eq.(\ref{tr}), (\ref{mphi}) and (\ref{gammaphi}):
\begin{equation}
 h\simeq \sqrt{\frac{16\pi}{m_\phi}}\left(\frac{g_*\pi^2}{10M_G^2}\right)^\frac{1}{4} \left(\frac{g}{v^2}\right)^{\frac{n-2}{n}} T_R.
\label{htr}
\end{equation} 
Then the baryon asymmetry ${n_B}/{s}$ is given in terms of $g$, $k$, and
$T_R$ by
\begin{equation}
  \frac{n_B}{s} \simeq 8.2 \times 10^{-11} \left(\frac{T_R}{10^6\mathrm{GeV}}\right)\left(\frac{2h}{n(n-1)g}\right)\left(\frac{m_{\nu_3}}{{0.05\mathrm{eV}}}\right)\frac{1}{\sin^2 \beta}\delta_{eff},
\label{tocompare}
\end{equation} 
where $h(g,k,T_R)$ is given by Eq.(\ref{htr}) with $v$ determined by
Eq.(\ref{v}) and (\ref{efold}).

\begin{figure}[tbp]
 \begin{center}
 \includegraphics[width=7cm]{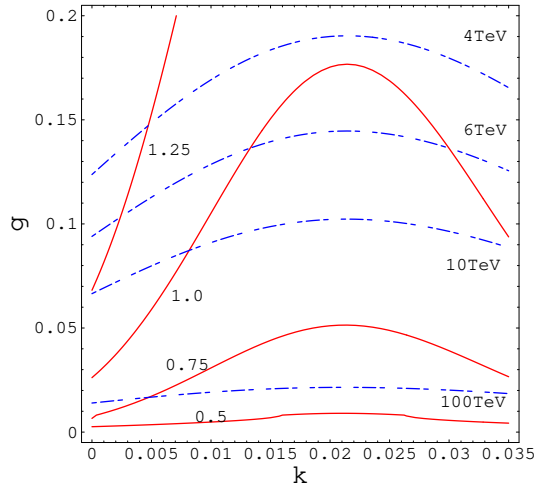}
  \caption{The contours of
 $\left(\frac{n_B}{s}\right)/\left(\frac{n_B}{s}\right)_{0}$ for $n=4$, $T_R=4 \times 10^6$GeV, $\delta_{eff}=1$, $\sin \beta=1$ are plotted in red (solid)
 lines. The
 blue (dashed-dotted) lines correspond to the contours of gravitino mass.}
 \label{bar}
 \end{center}
\end{figure}

In Fig.\ref{bar},
we plot the contours of $m_{3/2}$ and $(\frac{n_B}{s})/(\frac{n_B}{s})_0$ for
$T_R=4\times 10^6$GeV, $m_{\nu_3}=0.05$eV, $\delta_{eff}=1$, $\sin \beta=1$, where
$(n_B/s)_0$ is the baryon asymmetry of the universe suggested by WMAP
\cite{WMAP}:
\begin{equation}
 \left(\frac{n_B}{s}\right)_{0} \simeq 8.7 \times 10^{-11}.
\end{equation}
We note that the baryon asymmetry and the gravitino mass
for different reheating
temperatures can also be seen from Fig.\ref{bar}: The baryon asymmetry is
proportional to the square of the reheating temperature $T_R^2$, since the
coupling $h$ is approximately proportional to the reheating temperature.
As for the gravitino mass, its value is almost independent of $T_R$,
since $v$ is almost independent of $T_R$. 

This figure shows that the sufficient baryon asymmetry is produced in
a typical parameter region of the large-cutoff hypothesis:
$k\sim0.01$, $g\sim0.01-0.1$,
and $T_R\sim10^6$GeV, which turns out to be low enough to avoid the gravitino
overproduction. Thus it is revealed that the large-cutoff hypothesis is also
consistent with the observed baryon asymmetry.

\section{Discussion}\label{se:discussion}

We have studied the large-cutoff hypothesis from the viewpoint of
cosmology.
We first confirmed
that the spectral index in the new inflation
model has an upper bound $n_s\lsim 0.95$ (see Ref.\cite{Izaw})
and the large-cutoff hypothesis implies its boundary value,
which remarkably
agrees with the present experimental suggestion $n_s=0.95 \pm
0.02$ \cite{BOOMERANG}.
Secondly, we found a concrete setup where the sufficient baryon
asymmetry can be produced via non-thermal leptogenesis with the reheating
temperature low enough to avoid the gravitino overproduction in a typical
parameter region of large-cutoff hypothesis. 

We again emphasize
that the large cut-off hypothesis has several advantages
from the viewpoint of particle-physics phenomenology.
It solves the FCNC problem
and produces the mass spectrum $m_0\sim10 m_{1/2}\sim 10\mu$,
which yields the correct electroweak symmetry breaking
\cite{Ibe}.
Furthermore,
the spectrum realized in the large-cutoff hypothesis
accommodates the
appropriate amount of the dark matter density
\cite{Mor}.

We also mention CP violations in the visible-sector
supersymmetric standard model
as a sensitive low-energy probe of the supersymmetry breaking.
Phases of the theory would be limited severely if the scalar masses were
to be less than the TeV scale.
In contrast, for $m_{0}\sim10$TeV, such a constraint
is far milder, with the very heavy scalar masses
expected to be realized in the large-cutoff hypothesis
from the viewpoint of electroweak symmetry breaking and dark matter,
as mentioned above.

The heavy scalar masses are remarkably consistent
with the cosmological constraint, as we saw in this paper.
Thus we conclude that the large-cutoff theory with the supergravity new inflation
and non-thermal leptogenesis is consistent with all the phenomenological requirements from cosmology and particle physics.

Finally we comment on other types of inflations.
The presence of the large cutoff seems advantageous for
other inflationary models such as hybrid inflation and chaotic inflation.
In particular, large-field inflations imply the presence
of a larger scale
(see Ref.\cite{Feldstein:2005bm})
than the reduced Planck scale.
In fact, we suspect that multiple inflations may be so generic
as to include various types of inflations as their components,
whose slow-roll conditions are realized by the large-cutoff mechanism
\cite{Ibe}.

\section*{Acknowledgments}
M.~I. thanks the Japan Society for the Promotion of Science
for financial support. This work is partially supported by Grand-in-Aid Scientific Research (s) 14102004.

\appendix
\section*{Appendix: Another source of baryon and gravitino}\label{appe}

In section \ref{se:baryon}, we put aside the baryon asymmetry and the gravitino
produced through the coherent oscillation of right-handed sneutrino. Here we argue that this
contribution can be small enough to be neglected.

Firstly we explain the motion of right-handed sneutrino field which is the source of the
baryon asymmetry and gravitino. During inflation, the right-handed sneutrino is fixed at the origin due to the Hubble mass.
After the inflaton starts to roll down to the
vacuum, the mass of the right-handed sneutrino changes along the motion
trajectory of the
inflaton. As the oscillation energy of the inflaton decreases, the
origin of right-handed sneutrino becomes unstable, and right-handed sneutrino also starts
oscillation. Then the decay of right-handed sneutrino becomes
significant.\footnote{The decay
width of right-handed sneutrino is much larger than that of inflaton,
due to a large Yukawa coupling of right-handed neutrino
and standard-model particles compared with Eq.(\ref{hNN}).}
The baryon asymmetry and gravitino are provided by the
decay of this right-handed sneutrino
\cite{Murayama:1993em}.

Let us estimate the yields of the baryon asymmetry and gravitino provided
through the coherent oscillation of  right-handed sneutrino. As
mentioned above, the decay of right-handed sneutrino becomes significant
when the motion of right-handed sneutrino is induced by that of inflaton. 
Then the yields of the baryon asymmetry $n^N_B/s$
and the grvitino number $n^N_{3/2}/s$ produced
at the decay time of right-handed sneutrino are given by
\begin{eqnarray}
 \frac{n^N_B}{s}&\simeq&\varepsilon \frac{\rho_N}{m_N}\frac{45}{2\pi^2 g_* T_N^3} \\
 \frac{n^N_{3/2}}{s} &\simeq& Y^\phi_{3/2} \frac{T_N}{T_R}.
\end{eqnarray}
Here, $\varepsilon$ denotes the CP-asymmetry in right-handed sneutrino decay
defined in Ref.\cite{Fuk},
$T_N$ is the temperature of radiation
produced by right-handed sneutrino decay,
$Y^\phi_{3/2}$ is the yield of gravitino produced by inflaton
decay, and $\rho_N$ is the energy of the right-handed sneutrino at the
right-handed sneutrino decay. 

After the inflaton decay, these yields are diluted by the dilution
factor $\Delta$ estimated as
\begin{equation}
\Delta \simeq \frac{T_N}{T_R}\frac{\rho_\phi}{\rho_N},
\end{equation}
where $\rho_\phi$ is the energy of the inflaton at the right-handed
sneutrino decay.
Thus $n^N_B/s$ and $n^N_{3/2}/s$ after the inflaton decay are given by
\begin{eqnarray}
 \frac{n^N_B}{s}&\simeq&\varepsilon \frac{\rho_N}{m_N}\frac{45}{2\pi^2 g_* T_N^3} \frac{T_R}{T_N}\frac{\rho_N}{\rho_\phi}
  \simeq 5.3\times 10^{-11}\left(\frac{T_R}{10^6\mathrm{GeV}}\right)\left(\frac{m_{\nu_3}}{0.05\mathrm{eV}}\right)\frac{\rho_N}{\rho_\phi}\delta_{eff}\\
 \frac{n^N_{3/2}}{s} &\simeq& Y^\phi_{3/2} \frac{\rho_N}{\rho_\phi}.
\end{eqnarray}
These values are smaller than the yields produced at inflaton decay for
$\rho_N \ll \rho_\phi$
(see Eq.(\ref{tocompare})),
which we assume in the main text.

In fact, we checked that
$\rho_N \ll \rho_\phi$ is realized in a typical parameter region by solving
the equations of motion numerically for $n=4$.
We note a possibility that parametric resonance
occurs in specific points, and the energy
of right-handed sneutrino  $\rho_N$ becomes
comparable to that of inflaton  $\rho_\phi$
in such a case.

\end{document}